\documentclass[12pt]{iopart}
\usepackage{graphicx,cite,xcolor,float,iopams}
\expandafter\let\csname equation*\endcsname\relax
\expandafter\let\csname endequation*\endcsname\relax
\usepackage[pagebackref=false,colorlinks,linkcolor=blue,citecolor=blue,urlcolor=blue]{hyperref}
\usepackage{amsmath,amssymb}
\usepackage{blkarray,multirow}

\begin{document}
\title{Thermal transport and thermoelectric properties of transition metal dichalcogenides Mo$X_2$ from first-principles calculation}
\author{Radityo~Wisesa$^1$, Anugrah~Azhar$^{1,2}$, Edi~Suprayoga$^{3,\dagger}$}

\address{$^1$Physics Study Program, Syarif Hidayatullah State Islamic University Jakarta, South Tangerang 15412, Indonesia}
\address{$^2$Department of Physics and Astronomy, The University of Manchester, Oxford Road, Manchester M13 9PL, United Kingdom}
\address{$^3$Research Center for Quantum Physics, National Research and Innovation Agency (BRIN), South Tangerang 15214, Indonesia}
\ead{$^\dagger$edi.suprayoga@brin.go.id}

\begin{abstract}
  The properties of two-dimensional (2D) materials have been extensively studied and applied in various applications. Our interest is to theoretically investigate the thermal transport and thermoelectric properties of the 2D transition metal dichalcogenides Mo$X_2$ ($X$ = S, Se, Te). We employ density functional theory and Boltzmann transport theory with relaxation-time approximation to calculate the electronic and transport properties. We also implemented the kinetic-collective model to improve the calculation of lattice thermal conductivity. Our calculations indicate that MoTe$_2$ has the highest ZT of 2.77 among the other Mo$X_2$ at 550 K due to its low thermal conductivity and high electrical conductivity. Consequently, we suggest that Mo$X_2$ monolayers hold promise as materials for energy conversion devices due to their relatively high ZT. Moreover, these results could be beneficial to design 2D material based high performance thermoelectric devices.

\end{abstract}


\section{Introduction}
\label{sec:intro}
The discovery of graphene, a carbon-based material with a hexagonal structure that is only one atom thick, has generated great research interest in other two-dimensional (2D) materials. Research has shown that 2D materials have exotic properties that could be beneficial in future applications \cite{Fiori2014, Cooper2012, Choi2017}. For example, as the thickness of graphene increases, its electrical conductivity decreases, and it becomes more sensitive to temperature changes \cite{Fang2015}. Furthermore, 2D materials are used in a variety of applications, such as electronics, spintronics, energy storage, integrated memory, and energy conversion devices \cite{Wang2018, Splendiani2010, Ahn2020, Li2020}.

We are interested in the possibility of 2D materials in energy conversion devices, particularly thermoelectric (TE) devices. TE materials are materials that can convert heat into electricity. The performance of the TE material is measured by a dimensionless figure of merit ZT, which is determined by the electrical conductivity ($\sigma$), Seebeck coefficient ($S$), and thermal conductivity from the electronic contribution ($\kappa_e$), and phonon (or lattice) contribution ($\kappa_\mathrm{ph}$) according to the equation
\begin{equation}\label{ZT}
    \mathrm{ZT}=\frac{S^2\sigma}{\kappa_e+\kappa_\mathrm{ph}}T,
\end{equation}
where $T$ is the operating temperature. Theoretically, to have good thermoelectric performance, a material should possess a high Seebeck coefficient and electrical conductivity, as well as low thermal conductivity. Unfortunately, graphene has a low Seebeck coefficient and high thermal conductivity, which results in poor TE performance \cite{Xu2014}. Consequently, we must look for other materials. Monolayer transition-metal dichalcogenide (TMDC) is one of the 2D materials that are expected to have excellent TE properties \cite{Nakamura2018, C7TC01088E}.

\begin{figure}
    \centering
    \includegraphics[width=6.0cm]{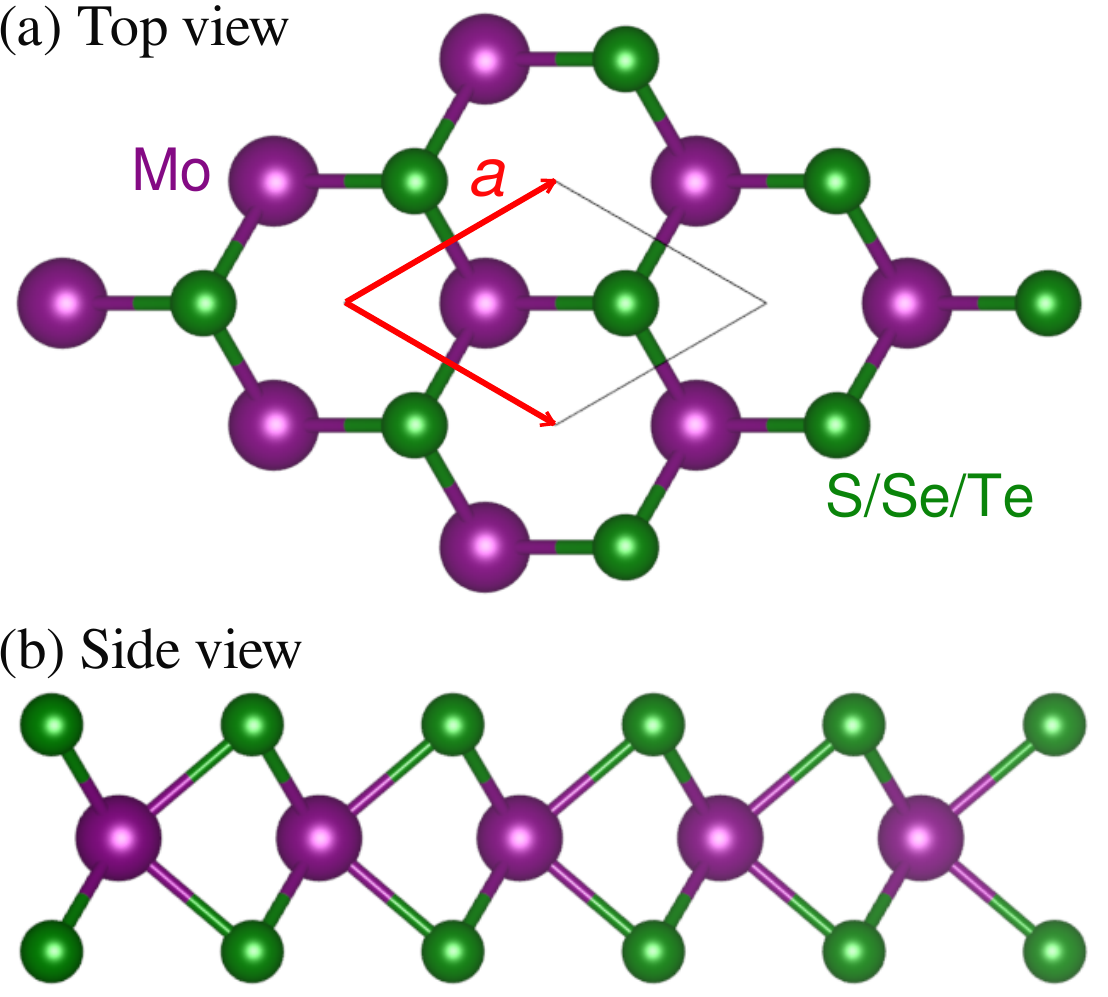}
    \caption{Geometrical structure of monolayers Mo$X_2$ with a hexagonal lattice.}
    \label{fig:structure}
\end{figure}

One of the interesting 2D TMDCs is the monolayer Mo$X_2$, where $X$ is sulfur, selenium, or tellurium. These materials have a honeycomb-like hexagonal structure consisting of a transition metal atom sandwiched between two chalcogenide atoms, as illustrated in Fig.~\ref{fig:structure}. To make a monolayer, we can use certain exfoliation techniques \cite{Jawaid2016, Zhang2020, Moh2018} to separate each bulk TMDC layer, which is held together by a weak van der Waals interaction. This fascinating TMDC structure has revealed some distinguishing properties. For example, bulk and few-layer MoS$_2$ exhibits an indirect bandgap, but monolayer MoS$_2$ exhibits a direct bandgap \cite{Mak2010}. As the number of layers is reduced, the electrical conductivity of the TMDC increases without affecting much of the Seebeck coefficient \cite{Hong2016}. Furthermore, monolayer Mo$X_2$ has been shown to have relatively low thermal conductivity \cite{Zulfiqar2019}. These properties make the monolayer Mo$X_2$ a promising material for thermoelectric applications.

In addition, various theoretical approaches have been utilized to assess the thermoelectric performance of 2D TMDCs \cite{TE1, TE2, TE3, TE4, TE5}. Recent studies have also shown that the examination of the strain, anisotropy, and electrical modulation of 2D TMDCs can improve their transport performance \cite{strain1, strain2, strain3, anisotropic1, anisotropic2, electrostatic}. However, currently there is no theoretical analysis that can accurately predict lattice thermal conductivity, in agreement with the experimental results. Consequently, a precise methodology for computing lattice thermal conductivity is imperative to assess the thermoelectric performance of 2D TMDCs.

In this study, we purpose to theoretically investigate the thermal transport and thermoelectric properties of a monolayer Mo$X_2$. We consider the electron relaxation-time approximation in the Boltzmann transport equation to improve the previous TE calculation results \cite{Wickramaratne2014}. We also implement the kinetic collective model in our calculation to explain previous experimental results of lattice thermal conductivity \cite{Torres2019}. 

\section{Computational Methods}
\label{sec:method}
Density functional theory calculation is performed using {\sc{Quantum Espresso}} (QE) \cite{Giannozzi2009} with ONCV pseudopotentials \cite{Hamann2013} based on the Perdew-Burke-Enzerhof (PBE) functional \cite{PBE}. To eliminate the interaction between adjacent layers, a vacuum layer of $20$ angstroms is added. In the self-consistent field (SCF) and relaxed structure calculations, we used a $80$ Ry cutoff energy and a $12\times12\times1$ $k$-point grid for the integration in the Brillouin zone. The optimized structures are shown in table \ref{tab:Egap}, consistent with the previous experimental results \cite{MoS2_a, MoSe2_a, MoTe2_a}. A finer $k$-point grid of $40\times40\times1$ is used to accurately calculate the electronic properties of each monolayer Mo$X_2$. In these systems, a non magnetic calculation with spin-orbit coupling (SOC) is employed to describe the electronic and transport properties of monolayer Mo$X_2$.

We compute the phonon dispersion of each material through density functional perturbation theory, as implemented in QE, with $20\times20\times1$ $k$-points and $8\times8\times1$ $q$-points grids. We then used {\sc{Phono3py}} code \cite{phono3py} to investigate the thermal transport properties of the monolayer Mo$X_2$. In this step, we used the $2\times2\times1$ supercell with displacements to calculate the phonon thermal conductivity using the linearized Boltzmann transport equation (LBTE) \cite{LBTE}. To improve the calculation, we consider the kinetic collective model (KCM) in monolayer Mo$X_2$. In this model, the phonon thermal conductivity is divided into two parts, a kinetic contribution ($\kappa_k$) and a collective contribution ($\kappa_c$), which are weighted by a switching factor $\Sigma\in[0,1]$ as expressed in the following equations \cite{kcm2014, Torres2017}:
\begin{equation}
    \kappa_\mathrm{ph} = (1-\Sigma) \kappa^*_k + \Sigma \kappa^*_c = \kappa_k + \kappa_c
    \label{eq:kph}
\end{equation}
\begin{equation}
    \kappa^*_k=\int\hbar\omega\frac{\partial f}{\partial T}\nu^2\tau_kDd\omega
\end{equation}
\begin{equation}
    \kappa^*_c=F\int\hbar\omega\frac{\partial f}{\partial T}\nu^2\tau_cDd\omega
\end{equation}
\begin{equation}
    \Sigma=\left (1+\frac{\tau_N}{\tau_R} \right )^{-1}
\end{equation}
where $F$ is the form factor determined by the sample geometry, $f(\omega,T)$ is the Bose-Einstein distribution function, $\nu(\omega)$ is the phonon mode velocity, and $D(\omega)$ is the phonon density of states as a function of phonon frequency, $\omega$. Here, $\tau_k$ and $\tau_c$ are the temperature-dependent relaxation times of phonons from a kinetic and collective contribution, respectively. The kinetic term corresponds to the effect of individual phonons, while the collective terms represent the overall contribution resulting from the coupling of phonon modes.
To consider 2D phonon thermal transport, the computed lattice thermal conductivity of each material should be normalized by multiplying it by the simulation box thickness divided by the monolayer width. Then, we compare our calculated lattice thermal conductivity with previous experimental results of the 2D-Mo$X_2$ lattice thermal conductivity.

Furthermore, to calculate the electronic transport coefficient of monolayers Mo$X_2$, we use the semi-classical Boltzmann transport equation (BTE) under the relaxation time approximation (RTA) as implemented in {\sc{BoltzTraP2}} \cite{Madsen2018}. In the BTE, the Seebeck coefficient, the electrical conductivity, and the thermal conductivity from the electrons contribution are defined as
\begin{equation}
	S = \frac{1}{eT} \frac{\mathcal{L}_1}{\mathcal{L}_0},\quad
        \sigma = \mathcal{L}_0,\quad
	\kappa_{e} = \frac{1}{e^2T} \left[\mathcal{L}_2 - \frac{\mathcal{L}_1^2}{\mathcal{L}_0} \right],
	\label{eq:seebeck}
\end{equation}
where $e$ is the electron charge and $\mathcal{L}_\alpha$ is the kernel of the generalized transport coefficient:
\begin{equation}
	\mathcal{L}_\alpha = e^2 \sum_{n,\textbf{k}} \tau_{n,\textbf{k}}v^2_{n,\textbf{k}}(\varepsilon_{n,\textbf{k}}- 
        \mu)^\alpha \left(-\frac{\partial f_{n,\textbf{k}}}{\partial \varepsilon_{n,\textbf{k}}} \right),\quad
        \alpha = 0,1,2.
	\label{eq:L}
\end{equation}
Here, $\tau_{n,\textbf{k}}$, $v_{n,\textbf{k}}$, $\varepsilon_{n,\textbf{k}}$, and $f_{n,\textbf{k}}$ respectively are the electronic relaxation time, group velocity, energy, and Fermi-Dirac distribution function of $n^{th}$ band at the wave vector $\textbf{k}$. To overcome $\sigma$ and $\kappa_e$ of 2D systems in {\sc{BoltzTraP2}}, again, we multiply those parameters by the thickness of the simulation box divided by the width of the monolayer. In the RTA calculations, the electron relaxation time is calculated by considering electron-phonon scattering, which can be defined as
\begin{equation}
    \frac{1}{\tau_{n,\textbf{k}}}=\frac{2~\mathrm{Im}(\Sigma_{n,\textbf{k}})}{\hbar},
    \label{eq:tau}
\end{equation}
where $\hbar$ is the reduced Planck constant and $\mathrm{Im}(\Sigma_{n,\textbf{k}})$ is the imaginary parts of the electron self-energy. In this work, we employ the {\sc{EPW}} code \cite{Ponce2016} to calculate the electron self-energy, using the same parameters as those used for the electronic and phonon structure calculations. Finally, we evaluate and compare the thermal transport and thermoelectric properties of the monolayers Mo$X_2$.

\section{Results and Discussion}
\label{sec:result}

\subsection{Electronic Structure}
\label{subsec:electronic}
\begin{figure*}
    \centering
    \includegraphics[width=15cm]{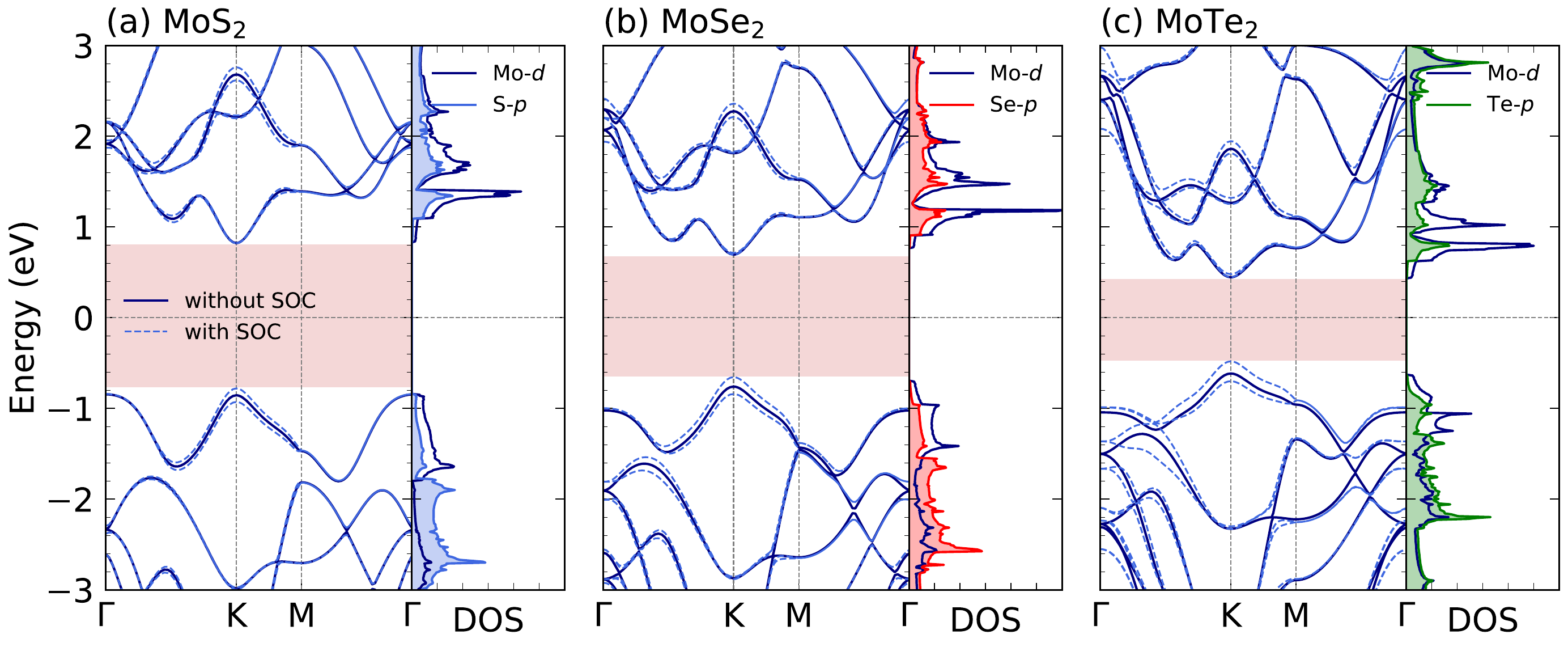}
    \caption{The electronic band structures and density of states of monolayer (a) MoS$_2$, (b) MoSe$_2$, and (c) MoTe$_2$. The solid and dashed lines show the band structure calculation without and with spin orbit coupling, respectively.}
    \label{fig:bands}
\end{figure*}

The thermoelectric properties of materials are mainly defined by their electronic structure. A good thermoelectric material should have an appropriate bandgap to maintain the electrical conductivity and Seebeck coefficient. Therefore, we inspect the electronic structure of each monolayer Mo$X_2$, which is shown in Fig.~\ref{fig:bands}. Our calculation shows that the three monolayers of Mo$X_2$ are direct semiconductors. 
We can see that the electronic bands are split by the SOC, which is generally larger in the valence band than in the conductor band. A large energy band splitting with a direct gap has been found at K points. Table \ref{tab:Egap} lists the bandgap values of three monolayers Mo$X_2$. Our calculation underestimates the bandgap value from prior experimental results \cite{Mak2010, Tonndorf2013, Ruppert2014}, as a typical result of the norm-conserving Vanderbilt pseudopotential.

\begin{table}
\caption{\label{tab:Egap}The calculated MoS$_2$, MoSe$_2$, and MoTe$_2$ lattice constant $a$ and bandgap energy $E_g$ with their previous experimental results.} 
\begin{indented}
\lineup
\item[]\begin{tabular}{@{}*{5}{l}}
\br                              
\multirow{2}{*}{Materials} & \multicolumn{2}{c}{$a$ (\AA)} & \multicolumn{2}{c}{$E_g$ (eV)}\\
    \cline{2-3} \cline{4-5}
    & Calculated & Experiment & Calculated & Experiment  \\
\mr
MoS$_2$\0 & 3.18 & 3.15 \cite{MoS2_a}  & 1.69 & 1.90 \cite{Mak2010} \cr
MoSe$_2$  & 3.32 & 3.30 \cite{MoSe2_a} & 1.46 & 1.57 \cite{Tonndorf2013, Roy2016} \cr 
MoTe$_2$  & 3.56 & 3.52 \cite{MoTe2_a} & 1.08 & 1.10 \cite{Ruppert2014, Roy2016} \cr 
\br
\end{tabular}
\end{indented}
\end{table}

Furthermore, our calculated density of states (DOS) for each material is consistent with the electronic band structure. All monolayers Mo$X_2$ carry a high and steep DOS, which implies a high Seebeck coefficient and can be a promising thermoelectric material, as previously demonstrated \cite{Pichanusakorn2010}. By performing strain engineering on Mo$X_2$ monolayers, we could take advantage of their DOS properties and achieve a higher and steeper DOS due to suppressed tunneling current \cite{Zhang2022}. Moreover, the projected density of states (PDOS) of Mo$X_2$ monolayers reveals that the Mo atom's $d$-orbital is the main contributor to the conduction bands, while the Mo atom's $d$-orbital and the $X$ atom's $p$-orbital dominate the valence bands. This indicates that the electronic properties of three monolayer Mo$X_2$ materials are determined by both the Mo atom and the $X$ atom.

\subsection{Phonon Properties and Lattice Thermal Conductivity}

\label{subsec:thermal}
\begin{figure*}
    \centering
    \includegraphics[width=13cm]{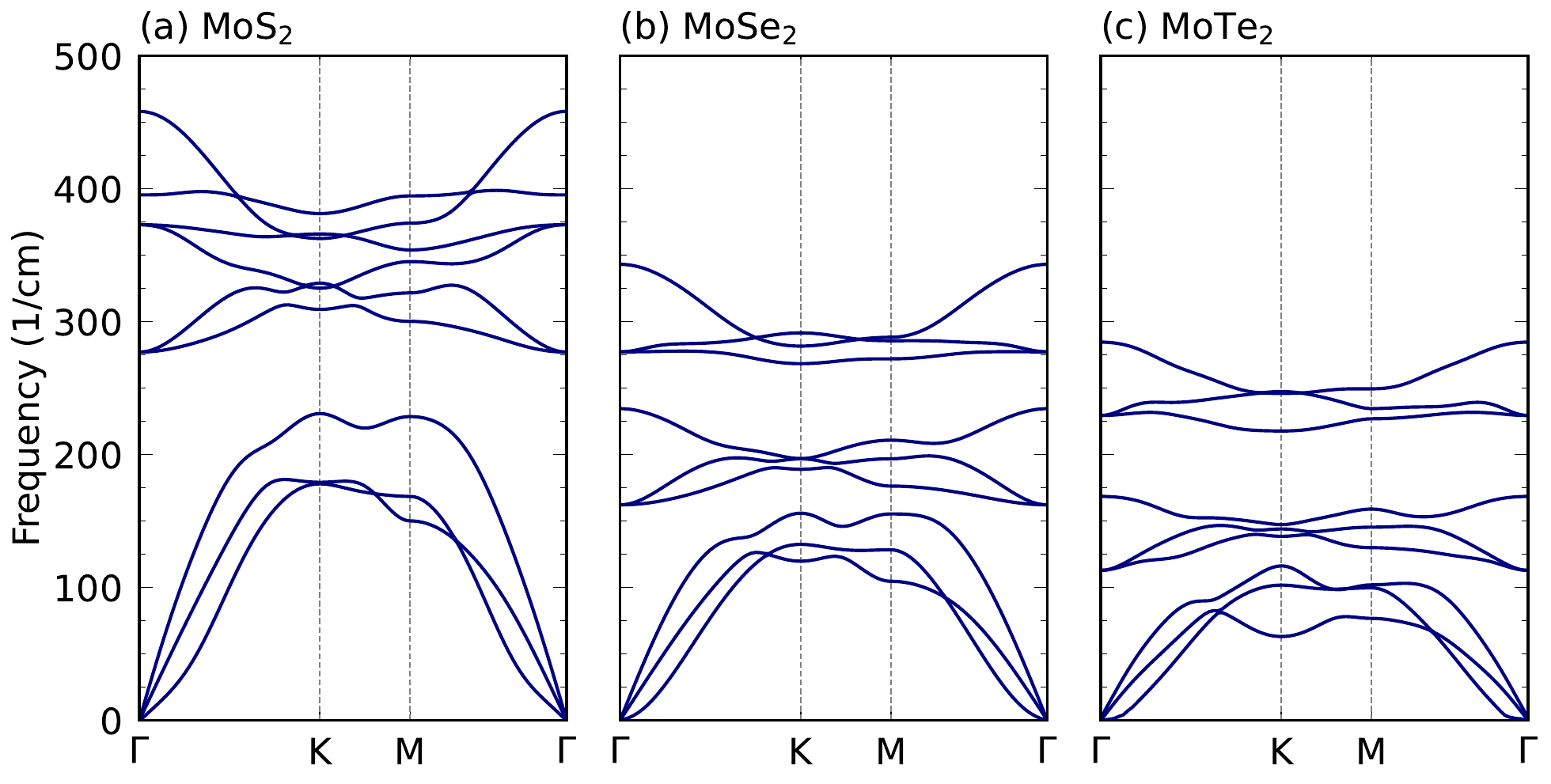}
    \caption{The phonon dispersion of monolayer (a) MoS$_2$, (b) MoSe$_2$ (b), and (c) MoTe$_2$. Non-negative frequency of the phonon indicating the stability of calculated structures.}
    \label{fig:phonon}
\end{figure*}

In order to gain a better understanding of the lattice thermal conductivity of a material, it is necessary to observe its lattice vibration behavior. Fig.~\ref{fig:phonon} shows the phonon dispersion along the $\Gamma$-K-M-$\Gamma$ high-symmetry line for all monolayer Mo$X_2$ materials. Three acoustic branches and nine optical phonons are present in the monolayer Mo$X_2$, which is made up of one Mo atom sandwiched between two $X$ atoms. The three branches of acoustic phonons are the transverse acoustic mode (TA), the longitudinal acoustic mode (LA), and the out-of-plane flexural acoustic mode (ZA). Our results indicate that MoTe$_2$ has the lowest phonon frequency, with a maximum frequency of approximately 285 cm$^{-1}$. This is due to the larger mass of the Te atom, which reduces its vibration. Moreover, the absence of imaginary vibrational frequency in our calculated phonon dispersion indicates that all monolayer Mo$X_2$ structures are dynamically stable.

\begin{figure*}
    \centering
    \includegraphics[width=15cm]{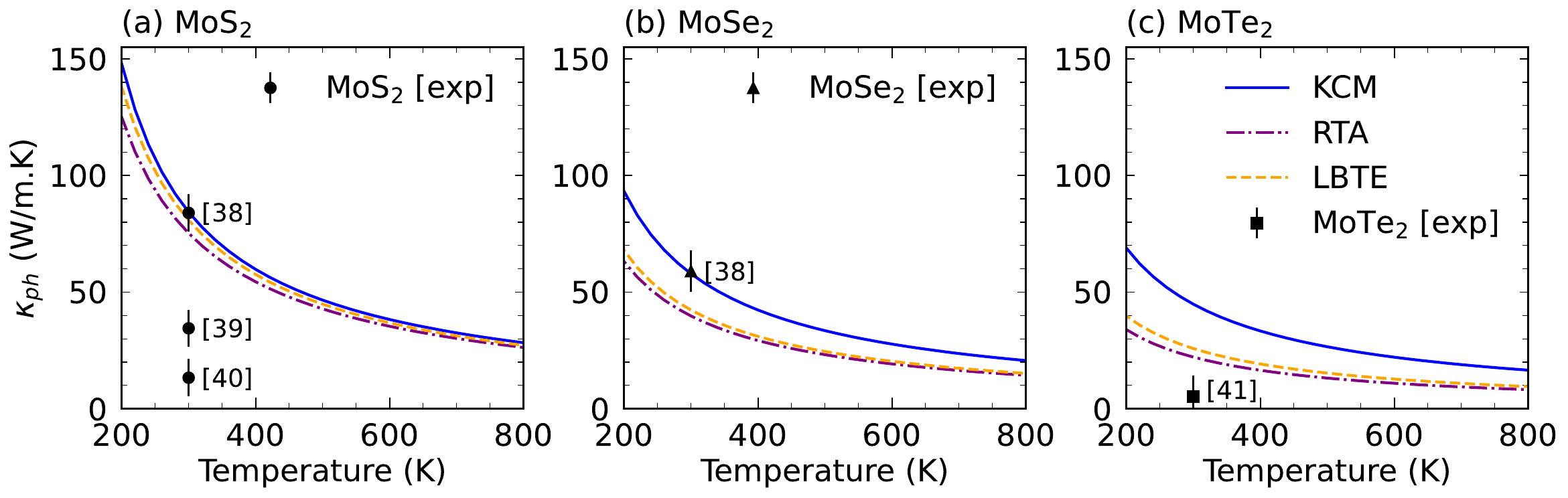}
    \caption{The computed lattice thermal conductivity of monolayer (a) MoS$_2$, (b) MoSe$_2$, and (c) MoTe$_2$ using KCM, RTA, and LBTE method. The experimental thermal conductivity was collected from the previous \cite{zhang2015measurement, Yan2014, C6NR09484H, Rodriguez2023}.}
    \label{fig:kph}
\end{figure*}

Fig.~\ref{fig:kph} shows the lattice thermal conductivity of monolayer Mo$X_2$ calculated using the KCM, RTA, and LBTE method. From the results, the KCM approach was found to be the most accurate, as it was consistent with previous experimental results\cite{zhang2015measurement}, as seen in Fig.~\ref{fig:kph} (a) and (b). This highlights the importance of the kinetic contribution of phonons in determining the lattice thermal conductivity of materials, particularly for monolayer Mo$X_2$. Furthermore, MoS$_2$ had the highest (84.3 W/m.K) and MoTe$_2$ had the lowest (43.3 W/m.K) lattice thermal conductivity of all monolayer Mo$X_2$. This is likely due to the tellurium atom, which has the greatest atomic mass among the sulfur or selenide atoms, which influences the frequency of its phonons and its lattice thermal conductivity.

\subsection{Thermoelectric Properties}
\label{subsec:tau}
\begin{figure*}
    \centering
    \includegraphics[width=15cm]{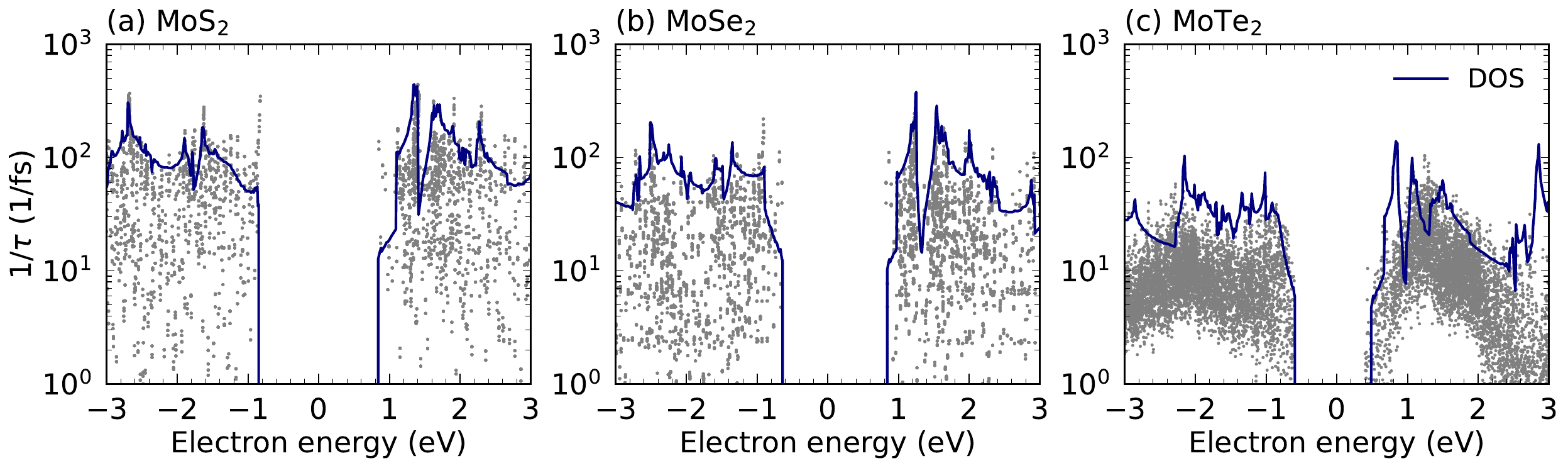}
    \caption{The electron relaxation time as a function of energy for each (a) MoS$_2$, (b) MoSe$_2$, and (c) MoTe$_2$ at 550 K.}
    \label{fig:tau}
\end{figure*}

Having the electronic structure of monolayer Mo$X_2$, we can calculate the Seebeck coefficient, the electrical and thermal conductivity using Eq.~\ref{eq:seebeck}. In this work, we use relaxation time approximation to solve the Boltzmann transport equation. We consider electron-phonon scattering to have a significant contribution in the electron mean-free-path, as defined in Eq.~\ref{eq:tau}. Fig.~\ref{fig:tau} shows the plot of each monolayer Mo$X_2$ computed electron relaxation time as a function of electron energy. It is clear that the inverse relaxation time is in tune with their electron density of states of each monolayer Mo$X_2$. Furthermore, our calculations are in agreement with previous theoretical studies \cite{Yadav2020}.

\begin{figure*}
    \centering
    \includegraphics[width=15cm]{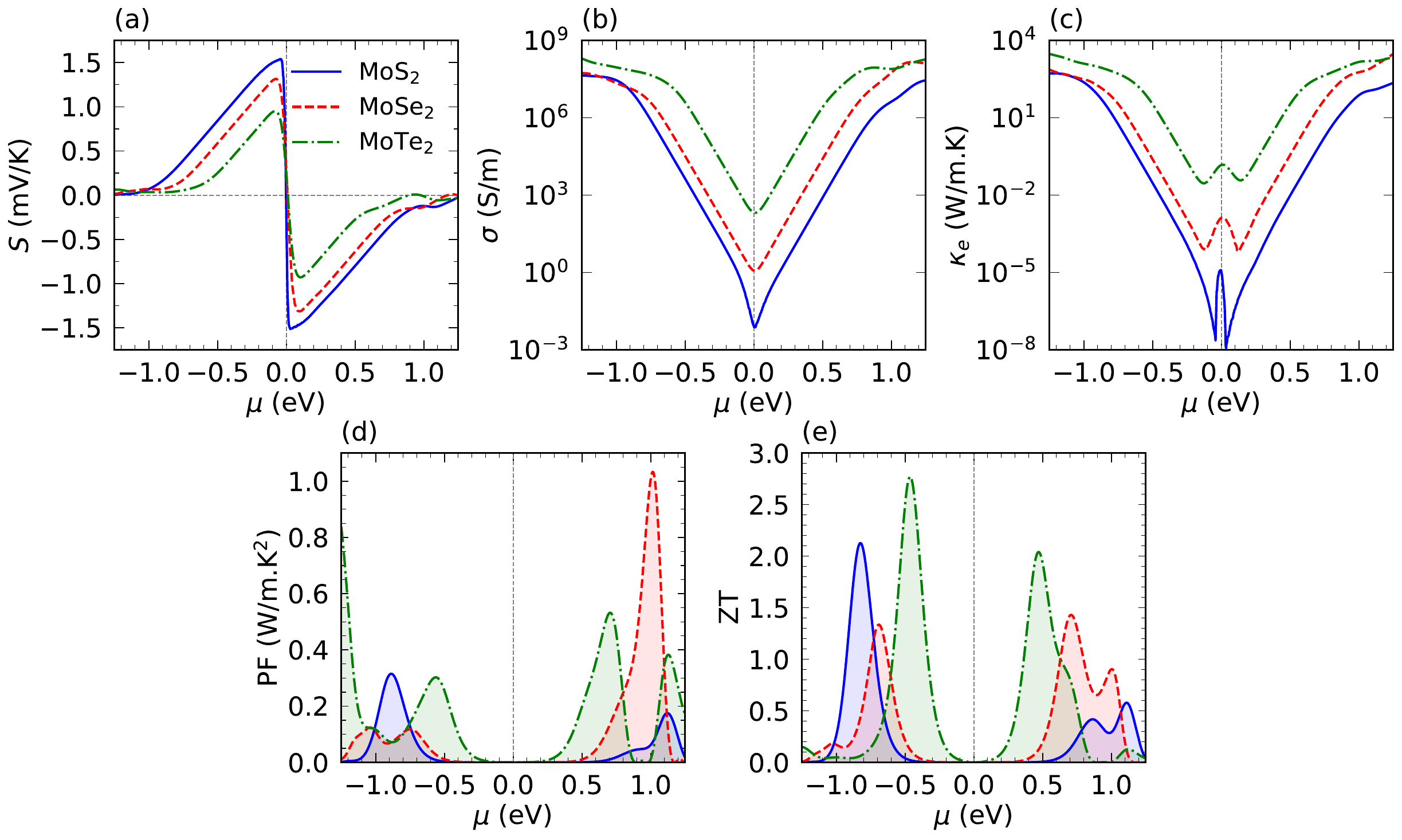}
    \caption{(a) Seebeck coefficient, (b) electrical conductivity, (c) electron thermal conductivity, (d) Power factor, and (e) ZT of MoS$_2$, MoSe$_2$, and MoTe$_2$ at 550K.}
    \label{fig:onsager}
\end{figure*}

Fig.~\ref{fig:onsager} summarizes the transport and thermoelectric properties of the monolayer Mo$X_2$. The highest Sebeeck coefficients are 1.54 W/mK for MoS$_2$, 1.32 W/mK for MoSe$_2$, and 0.95 W/mK for MoTe$_2$. Our calculations indicate that MoS$_2$ has the highest Seebeck coefficient, while MoTe$_2$ has the lowest. Furthermore, due to its narrow bandgap, MoTe$_2$ has the highest electrical conductivity, while MoSe$_2$ has the lowest. The electronic contribution of thermal conductivity in monolayer Mo$X_2$ has the same nature as the electrical conductivity, implying that the Wiedemann-Franz law is satisfied in these systems. At 550 K, the lattice thermal conductivity plays a more significant role in determining the thermal conductivity as a result of its higher value than the electronic thermal conductivity. Our study reveals that the monolayer MoTe$_2$ has the highest thermoelectric figure of merit (ZT) of 2.77 among the other monolayer Mo$X_2$ (2.13 for MoS$_2$ and 1.34 for MoSe$_2$). This is unexpected considering the low Seebeck coefficient and the thermoelectric power factor of MoTe$_2$. Despite having the lowest Seebeck coefficient, its high electronic conductivity and very low lattice thermal conductivity help to improve its ZT.

\begin{figure*}
    \centering
    \includegraphics[width=15cm]{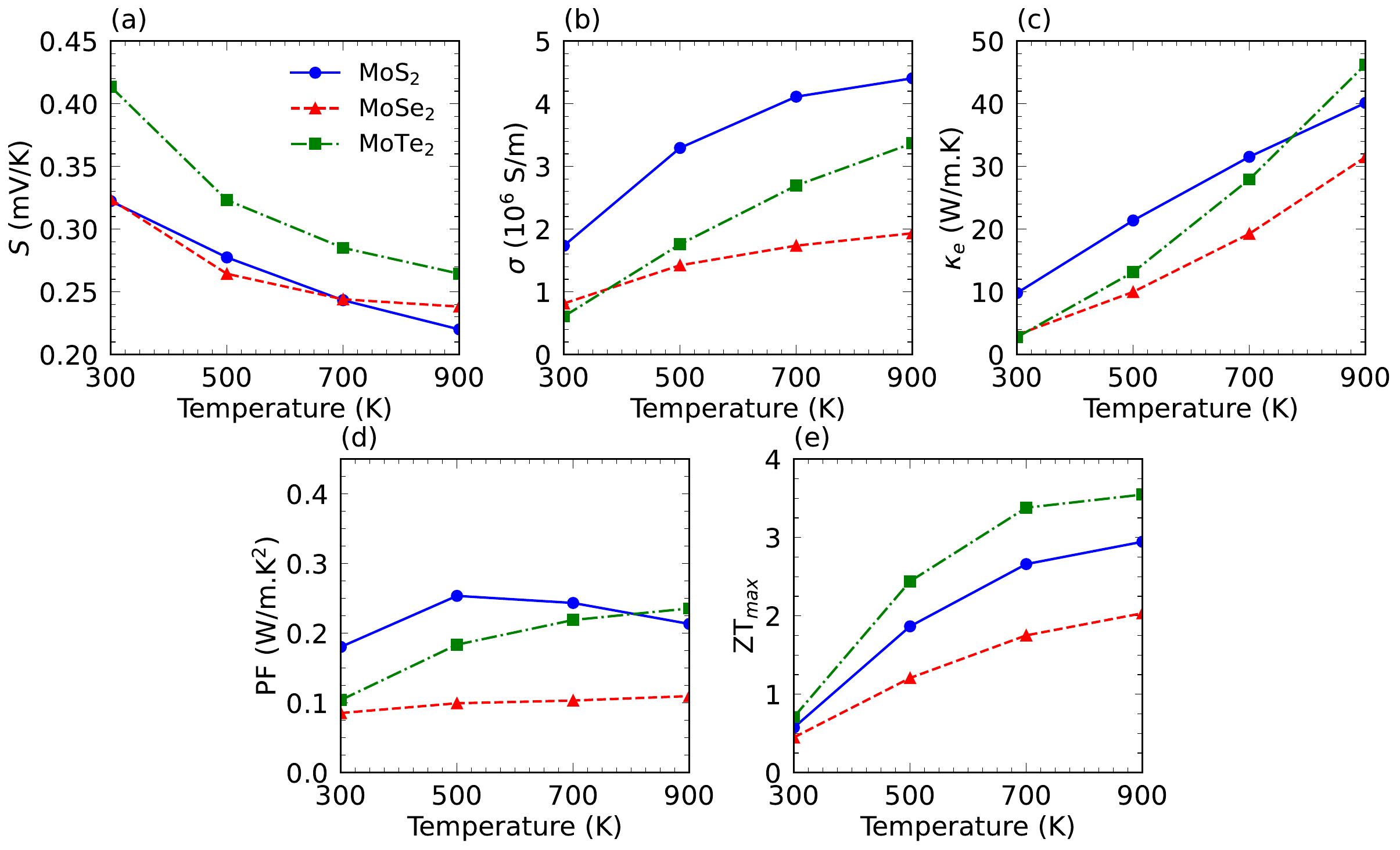}
    \caption{Temperature-dependent (a) Seebeck coefficient, (b) electrical conductivity, (c) electron thermal conductivity, (d) power factor, and (e) ZT of monolayer MoS$_2$, MoSe$_2$, and MoTe$_2$. The fixed energy level has been chosen at the point where the ZT exhibits the highest value for each material.}
    \label{fig:onsagger(t)}
\end{figure*}

This work also conducted the temperature-dependent thermoelectric properties of the monolayer Mo$X_2$. As seen in Fig.~\ref{fig:onsagger(t)}, the electrical and electronic thermal conductivities increase with temperature, despite the fact that the calculated scattering rates of monolayer Mo$X_2$ also increase with temperature, which can limit carrier mobility. Furthermore, the lattice thermal conductivity is an important factor in defining the total thermal conductivity of the monolayer Mo$X_2$ at lower temperatures. However, because of its high value at higher temperatures, the electronic thermal conductivity has become a more important parameter in determining the total thermal conductivity.

Fig.~\ref{fig:onsagger(t)} shows our calculation of the temperature-dependent thermoelectric power factor and ZT of the monolayer Mo$X_2$. Fig.~\ref{fig:onsagger(t)}(d) shows that the thermoelectric power factor of MoS$_2$ is highest at lower temperatures, while that of MoSe$_2$ is highest at higher temperatures due to the increase in the Seebeck coefficient at higher temperatures. According to our calculations, the ZT improves with increasing temperature (Fig.~\ref{fig:onsagger(t)}(e)). However, we can see that the ZT graph becomes sloping for monolayer Mo$X_2$. This is due to the increasing electronic thermal conductivity of each material. According to our findings, MoTe$_2$ has an extremely high ZT of 3.55 at a temperature of 900 K.

\section{Conclusions}
\label{sec:conclusion}
We have theoretically investigated the electronic structure and thermoelectric performance of  monolayer TMDC Mo$X_2$, where $X$ is the S, Se, and Te atoms.
Among the Mo$X_2$ monolayers, MoTe$_2$ has the lowest vibrational phonon frequency. This observation implies that MoTe$_2$ has a low phonon thermal conductivity. Using the KCM method, we found that MoTe$_2$ has an extremely low phonon thermal conductivity of approximately 43.3 W/mK at room temperature, as predicted.
Moreover, our calculation of the transport coefficient using the semi-classical Boltzmann transport equation and the relaxation time approximation shows that MoS$_2$ has the highest Seebeck coefficient, while MoTe$_2$ has the lowest. However, MoTe$_2$ has the highest electrical and electron thermal conductivity, while MoS$_2$ has the lowest. Furthermore, our calculations show that MoSe$_2$ has the highest thermoelectric power factor due to its Seebeck coefficient and electrical and electronic thermal conductivity. Surprisingly, having extremely low phonon thermal conductivity and high electronic conductivity, MoTe$_2$ exhibits the highest ZT among the other Mo$X_2$ materials with a ZT of 2.77. Our results also show an increase in ZT with an increase in temperature.

Finally, our results suggest that MoX$_2$ monolayers have considerable potential for use in energy conversion devices, especially in wearable devices, due to their compact size. In addition, these results could serve as a basis for further manufacturing and modifying the crystalline and electronic structures of MoX$_2$ monolayers to improve their thermoelectric performance. Therefore, we hope that our papers could provide insight for theoretical and experimental researchers in future studies.

\section*{Acknowledgements}
The numerical calculation was performed using the BRIN--HPC facilities. E.S. acknowledges support from the e-ASIA Joint Research Program (e-ASIA JRP).

\section*{References}
\bibliographystyle{iopart-num}

\end{document}